\documentclass[conference]{IEEEtran}
\IEEEoverridecommandlockouts
\usepackage{cite}

\ifCLASSINFOpdf
\usepackage[pdftex]{graphicx}
\else
\fi
\usepackage{url}

\begin{document}
%
\title{At Your Service: \\Coffee Beans Recommendation \\From a Robot Assistant}

\author{
    \IEEEauthorblockN{Jacopo de Berardinis\IEEEauthorrefmark{1}\thanks{Corresponding Email: jacopo.deberardinis@manchester.ac.uk. The authors Jacopo de Berardinis, Gabriella Pizzuto, Francesco Lanza and Jorge Meira contributed equally.}, Gabriella Pizzuto\IEEEauthorrefmark{2}, Francesco Lanza\IEEEauthorrefmark{3}, Antonio Chella \IEEEauthorrefmark{3}, Jorge Meira \IEEEauthorrefmark{4}, Angelo Cangelosi\IEEEauthorrefmark{1}}
    \IEEEauthorblockA{\IEEEauthorrefmark{1}School of Engineering, The University of Manchester}
    \IEEEauthorblockA{\IEEEauthorrefmark{2}School of Informatics, The University of Edinburgh}
    \IEEEauthorblockA{\IEEEauthorrefmark{3}Department of Engineering, University of Palermo}
    \IEEEauthorblockA{\IEEEauthorrefmark{4}Department of Computer Science and Information Technologies, University of A Coruña}
}



%


\maketitle

\begin{abstract}
With advances in the field of machine learning, precisely algorithms for recommendation systems, robot assistants are envisioned to become more present in the hospitality industry. Additionally, the COVID-19 pandemic has also highlighted the need to have more service robots in our everyday lives, to minimise the risk of human to-human transmission. One such example would be coffee shops, which have become intrinsic to our everyday lives. However, serving an excellent cup of coffee is not a trivial feat as a coffee blend typically comprises rich aromas, indulgent and unique flavours and a lingering aftertaste. Our work addresses this by proposing a computational model which recommends optimal coffee beans resulting from the user's preferences. Specifically, given a set of coffee bean properties (objective features), we apply different supervised learning techniques to predict coffee qualities (subjective features). We then consider an unsupervised learning method to analyse the relationship between coffee beans in the subjective feature space. Evaluated on a real coffee beans dataset based on digitised reviews,
our results illustrate that the proposed computational model gives up to 92.7 percent recommendation accuracy for coffee beans prediction. From this, we propose how this computational model can be deployed on a service robot to reliably predict customers' coffee bean preferences, starting from the user inputting their coffee preferences to the robot recommending the coffee beans that best meet the user's likings.
\end{abstract}

%
\IEEEpeerreviewmaketitle

\section{Introduction}

\begin{figure*}
    \centering
    \includegraphics[width=0.8\textwidth]{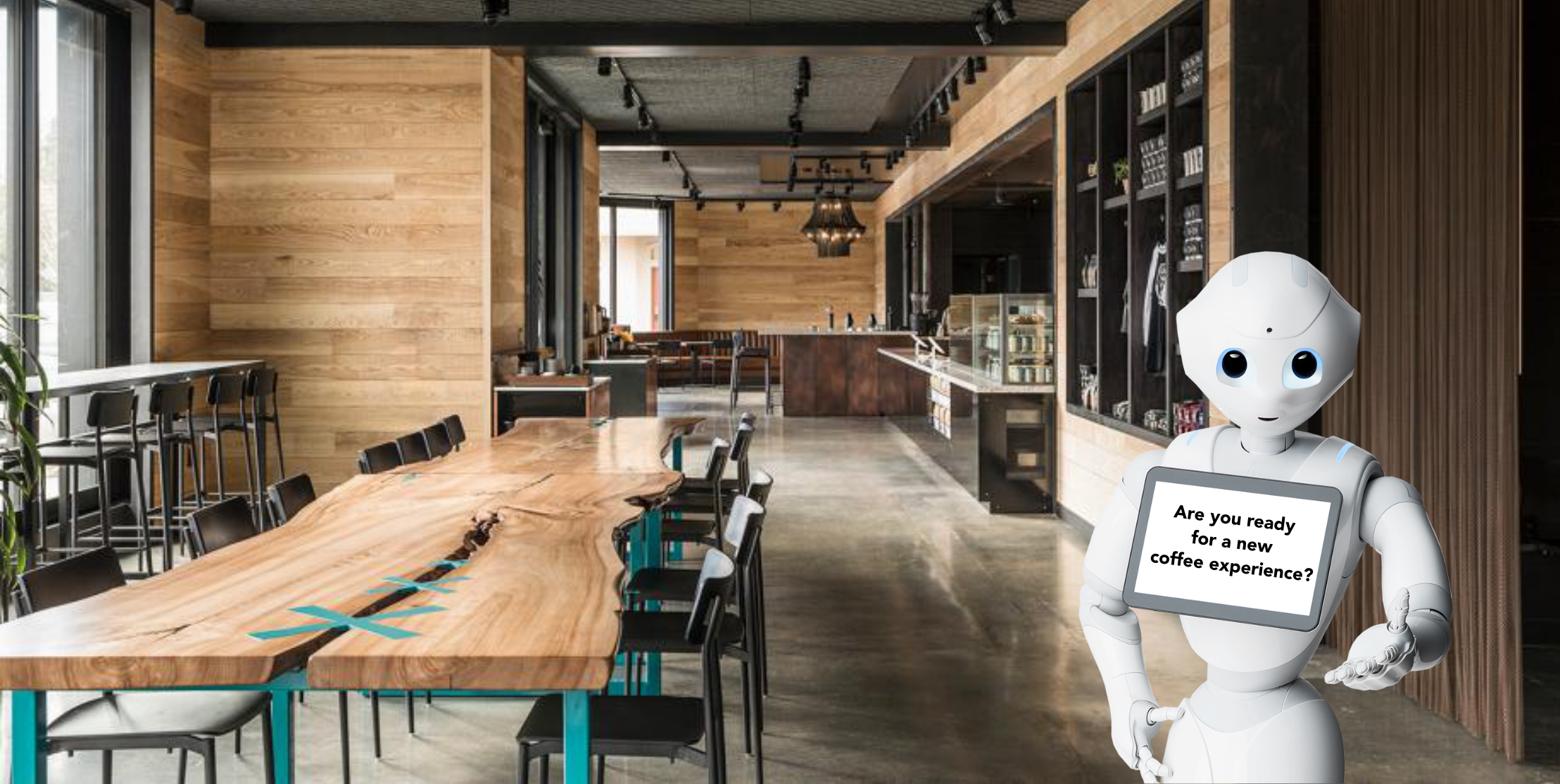}
    \caption{An illustration of how our coffee bean recommendation system would be deployed in a coffee shop. The humanoid robot, Pepper, invites the user to input his or her coffee preferences. Using our proposed recommendation system, Pepper suggests the perfect coffee blend for the customer.}
    \label{fig:teaser} 
\end{figure*}

The field of artificial intelligence has received increasing attention as it has the potential to improve different domains, for example the hospitality industry\cite{bowen2018beware,cain2019sci}. As robots are envisioned to become more present in our everyday lives, there is a growing interest to use service robots. Also, as we started realising with the COVID-19 pandemic, there is an accelerated need of service robots to minimise human to human transmission in such scenarios \cite{yang2020combating}. Service robots are becoming more present. Precisely, service robots should be capable of adapting to users' preferences, by understanding their needs to provide a more personal interaction \cite{fu2020sharing}. For example, robots in restaurants, or specifically coffee shops, would greatly enhance customer satisfaction with a more personalised experience \cite{nakanishi2020}. Witnessing this, an important challenge to address would be the deployment of an intelligent coffee recommender system\footnote{In this paper, we use the terms recommender and recommendation systems interchangeably.} on a humanoid robot in a coffee shop, as illustrated in Figure~\ref{fig:teaser}.

Recommendation systems are increasingly being used \cite{portugal2018}. In the hospitality industry, recommendation systems are being used to recommend healthy food options \cite{trangtran2018} or the perfect bottle of wine \cite{cruz2018}. Their strengths lie in their ability to perceive customers' food desires and recommend suggestions from computational analysis of users' interaction and related past information. Specifically, an intelligent recommender system builds a personalised experience when given specific characteristics. For example, in a coffee shop, the human-centred recommendation of coffee beans would rely on coffee bean qualities (subjective features) and coffee bean properties (objective characteristics). 

Although recommender systems have gained popularity in the hospitality domain \cite{trangtran2018}, \cite{parsons2009}, to the best of our knowledge, no coffee recommender system exists, and even more so, one which can be integrated with a robot. Hence, we aim to build a coffee recommendation system which suggests beans when the user expresses his or her subjective preferences. In this paper, we contribute to this by formulating the coffee bean recommendation problem as a multi-target regression problem. In future work, we plan on carrying human-agent interaction experiments and here we give an overview of how the overall system would work on a humanoid robot and propose the integration of this for a coffee recommender system. Beyond the recommendation itself, other factors (such as the modality) influence the user's overall experience. We envision that the use of an embodied agent to deliver the coffee bean choice would enhance the overall perception of the recommendation.

In this paper, we present a computational model for a coffee bean recommendation engine (Section~\ref{sec:methods}), extending our work presented in \cite{poster2020hai}. Precisely, we show how given a list of objective features (species, country of origin, region, variety, colour, category one and two defects, processing method, moisture) we can predict subjective traits (uniformity, sweetness, balance, body, flavour, acidity, aftertaste, aroma) using regression methods such as random forests (RFs), support vector regression (SVR) and multi-layer perceptron (MLP), to automate the coffee review process. Then, we also show that by using the nearest neighbour method, input data given by users can be compared to both predicted and available subjective components to recommend the most favourable coffee. Our computational model is evaluated on the Coffee Quality Institute database, comprising digitised reviews from arabica and robusta coffee beans (Section~\ref{sec:experimentalevaluation}). Given this computational model, we explain qualitatively how this would be embodied on a real robot in a coffee shop to improve overall experience and provide a more tailored user experience.

\begin{figure*}[t]
    \centering
    \includegraphics[width=0.99\linewidth]{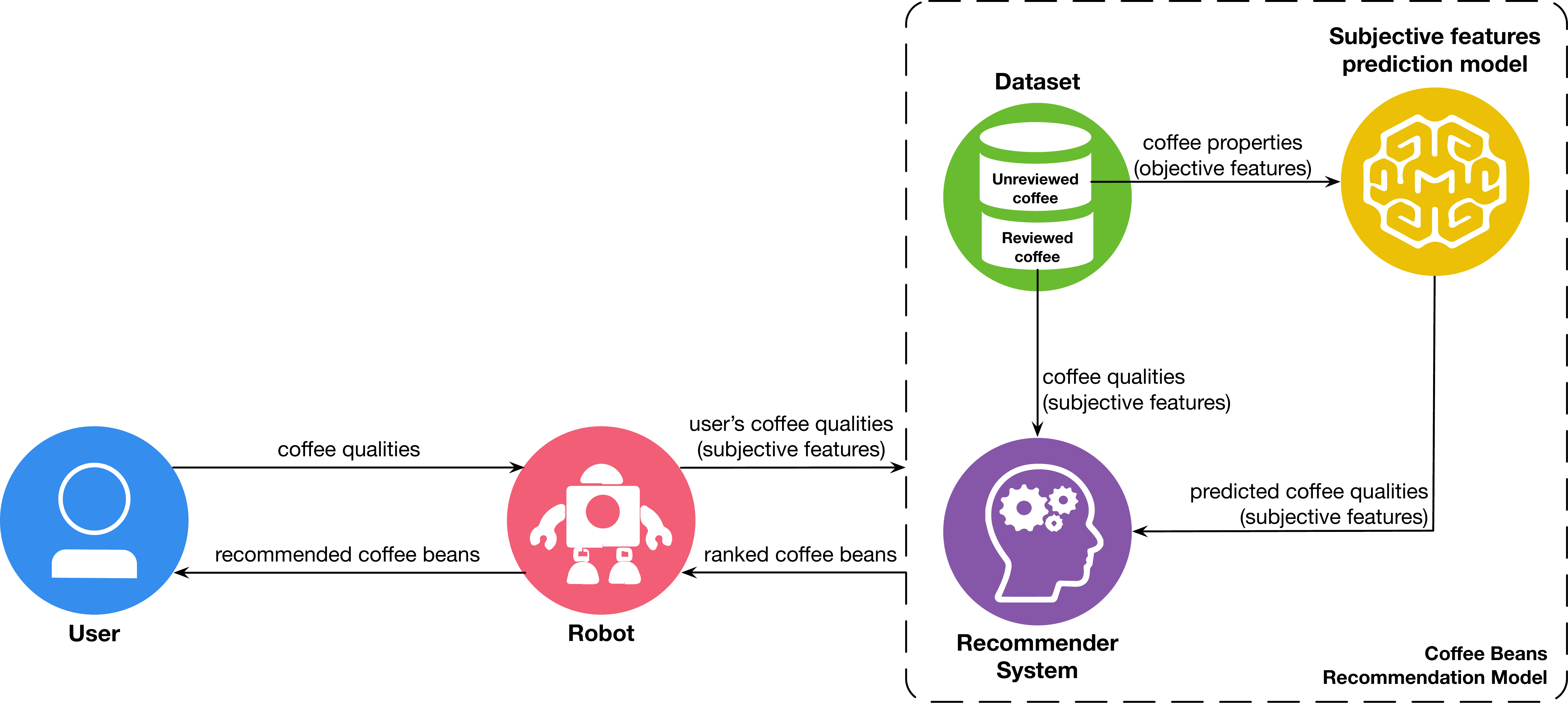}
    \caption{Illustration of the coffee bean recommendation system. Objective features from the dataset are fed into a regression model to predict subjective traits. The predictions, together with user input fed into the robot's tablet, are compared to obtain the coffee suggestion. In turn, this is given by the robot.}
    \label{overallblockdiagram}
\end{figure*}

\section{Related Work}
\label{sec:relatedwork}

There has been several research efforts to improve user experience in gastronomy, both from the machine learning and robotics communities. Since coffee is such a staple beverage in our everyday lives and it has been shown that coffee provides several health benefits \cite{bae2014}, numerous attempts have been made to improve its \textit{quality}. Goyal et al. illustrated how an artificial neural network (ANN) model can predict shelf life of roasted coffee flavoured drinks, given data samples from 50 observations (colour, appearance, flavour, viscosity, sediment) \cite{goyal2013}. The best result achieved was with a MLP with a single hidden layer of 12 neurons. Ribeiro et al. used partial least squares regression to show how the chemometric analysis of coffee beverage sensory data with the respective chromatographic profiles of volatile roasted coffee compounds can predict acidity, bitterness, flavour, cleanliness, body, and overall quality \cite{ribeiro2012}. Their results were congruent with the mean score of human coffee experts. The data used for our work is more similar to the one by Goyal et al. \cite{goyal2013} rather than by Ribeiro et al. \cite{ribeiro2012} due to the dataset we had available.

The quality of coffee can also be improved by using computer vision techniques. This field has seen remarkable progress with the introduction of neural networks, specifically to address the challenge of classifying coffee beans' roast level. Different roast levels, from light to dark, give different preservation of the coffee bean's natural aroma and flavour. As a result, there has been efforts to standardise this process. One of the works used a backpropagation neural network and obtained an accuracy of 97.5\% \cite{nasution2017}. The colour of coffee beans also illustrates their roast level. Morais de Oliveira et al. have focused on classifying coffee beans based on their colour. They combined ANNs with a Bayes classifier to reach 100\% accuracy in classifying the beans into four colour groups \cite{deoliveira2016}. Our research effort differs from these as we do not use machine learning models to monitor the quality of the coffee beans, but more as a tool for recommendation.

Recommender systems have received an increasing surge of attention, due to their attractive benefits of assisting the user during decision making when overloaded with information \cite{ricci2010}. Considering how important food choices are at this day and age, it is remarkable that only a small number efforts have been put towards deploying more gastronomy-related recommender systems. Tran et al. \cite{trangtran2018} give an overview of different systems and methods for recommending healthy food. The methods ranged from collaborative filtering to content and constraint-based recommendations. These systems focus on adapting recommendations to individuals, when considering users' food preferences and nutritional needs, or alternatively focus more on healthy contributions for group scenarios. With respect to beverage recommendation, most efforts have revolved around recommending wine \cite{parsons2009}. Although this is beneficial to consumers, considering the similarities in their respective smells and flavours \cite{croijmans2016}, coffee recommender systems are vital to aide consumers in their choices. Our work aims to address this open gap.

When it comes to robot deployment, current systems adapt the behaviour of the agent towards users' preferences through crowdsourcing and collaborative filtering to learn and predict task preferences, such as prioritising tidying-up tasks when given a scene \cite{abdo2015}. Service robots are also being used as baristas for remembering previous orders \cite{irfan2020} or to show neutral, entertaining, and emphatic behaviours when suggesting drinks and taking orders from customers \cite{rossi2020}. There have also been research efforts to develop a service robot in a bar setting \cite{giuliani2013}, \cite{foster2014} depicting social behaviours in response to states, affects, and when dealing with uncertainty of the customers during interaction. However, these works fail to address the personalisation problem from a multi-user perspective and in reality the user experience would be greatly improved if the robot relies on a recommender system that also takes into account users with similar preferences.   

In this paper, we bridge insights from these fields to address the real-world challenge of personalised coffee bean recommendation. Our work thus mainly focuses on how machine learning can be used to recommend coffee beans and how a robot would be used to deploy the system in the real world.

\section{Methodology}
\label{sec:methods}

Our work addresses the challenge of tailoring the choice of coffee to the user's preferences. To do so, we propose a computational model (Section~\ref{ssec:coffeemodel}) which relies on supervised and unsupervised learning techniques.
An overview of our solution is illustrated in Figure~\ref{overallblockdiagram}.
The user inputs his or her coffee likings through the tablet on the robot, which uses the recommender system to return the coffee suggestion.
In Section~\ref{ssec:introbot}, we explain how we implemented this idea, the important role which the robot plays in such a scenario, and how we would deploy this in the real world.

\subsection{Coffee beans recommendation model}
\label{ssec:coffeemodel}

Given that the coffee review process is expensive and requires expert domain knowledge, it is essential to consider that a portion of the coffee beans available in the market at a certain time might not yet have been reviewed.
The unavailability of official reviews is indeed a practical concern in the coffee industry, as it occurs every time a novel blend is introduced or in case of local and niche products.
Nevertheless, with the goal of providing the user with the best experience, a recommender system should also be able to suggest unreviewed coffee in case the expected coffee qualities (those that would result from an actual review process) match the user's preferences.
With these premises, we aim for a system supporting this functionality in the context of an informative and authoritative dataset that we can use for coffee recommendation.

\begin{figure}[h]
\centering
     \includegraphics[width=\linewidth]{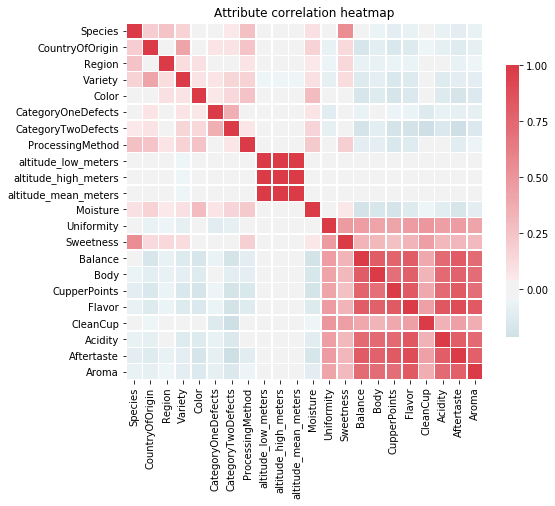}
     \caption{Pearson's correlation between the attributes considered during the last step of feature selection.}
     \label{attributecorrelationheatmap}
 \end{figure}

\subsubsection{Dataset and pre-processing}
The Coffee Quality Institute (CQI) is a nonprofit organization working internationally to improve the quality of coffee and the lives of the people who produce it. CQI provides training and technical assistance to coffee producers and other individuals in the supply chain to increase the value, volume and sustainability of high quality coffee production. CQI also works toward building institutional capacity in coffee producing countries by creating systems and infrastructure that encourage a focus on quality that leads to higher farmer incomes.

The dataset considered for of our system is based on the \textit{Coffee Quality Institute} database\footnote{All credits to Coffee Quality Institute for the data.}.
More precisely, a previously scraped version of the coffee quality dataset added to this repository\footnote{Link to the repository: \url{https://github.com/jldbc/coffee-quality-database}} was used.
This dataset comprises 1340 digitised reviews from 1312 arabica and 28 robusta coffee beans.
For each review, the dataset shows 44 features divided into \emph{quality measures} given by reviewers on a \emph{1-10 scale}, \emph{beans metadata} and \emph{farms metadata}.
The 44 features were cleaned and divided into \emph{objective} and \emph{subjective} features.

\begin{table}[h]
\centering
 \caption{Objective features (coffee properties) and subjective features (coffee qualities) after feature selection.}
 \label{tab:features}
 \begin{tabular}{|c|c|}
 \hline
 {\textbf{Objective Features}} & {\textbf{Subjective Features}} \\ \hline
  Species & Aroma  \\ \hline
  Country of Origin & Flavour \\ \hline
  Region & Body \\ \hline
  Variety & Sweetness  \\ \hline
  Color & Acidity \\ \hline
  Category One Defects & Balance \\ \hline
  Category Two Defects & Uniformity  \\ \hline
  Processing Method & Aftertaste \\ \hline
  Moisture &  \\ \hline
  \end{tabular}
\end{table}

Two feature selection strategies were adopted: (i) \textit{univariate feature selection}, identifying the most optimal features using univariate statistical tests; (ii) \textit{tree-based feature selection} to compute feature importance.
From our analysis, the altitude attributes were discarded as they did not contribute significantly to the regression model.
Finally, by inspecting the correlation between the selected features (Figure~\ref{attributecorrelationheatmap}), we ended up with 17 attributes.
These are species, country of origin, region, variety, color, category one defects, category two defects, processing method, moisture illustrated in Table \ref{tab:features}, with their characterisation into 
\textit{objective features} (coffee properties) and aroma, flavour, body, sweetness, acidity, balance, uniformity, aftertaste for \textit{subjective features} (coffee qualities).

For each feature, we give a brief description to understand the dataset, as shown in Table~\ref{tab:features}, in objective feature we have: 
\begin{itemize}
     \item \emph{Species} indicates the kind of the coffee beans. The values of this feature are \emph{arabica} or \emph{robusta};
     \item \emph{CountryOfOrigin} indicates the country of origin of beans;
     \item \emph{Region} indicates the region of the country of origin;
     \item \emph{Variety} indicates the variety of coffee. 
     \item \emph{Color} indicates the color of the greens.
     \item \emph{CategoryOneDefects} and \emph{CategoryTwoDefects} indicates the major and the minor defects in the beans.
     \item \emph{ProcessingMethod} indicates the method whereby the beans were handled and processed;
     \item \emph{Moisture} indicates the moisture percentage given by a green analysis.
\end{itemize}

Instead, for subjective features the system calculates scores, as human reviewers do, to quantify in a range of 1-10 a quality measure for each feature considered. The terms to convey subjective features are self-explanatory.

\subsubsection{Subjective features prediction and recommender system}

To simulate the realistic scenario mentioned before, we randomly select a subset of observations from the dataset and assume their subjective features to be unknown.
As illustrated in Figure~\ref{overallblockdiagram}, this partition is thus intended to represent unreviewed coffee.
We use regression methods to predict the subjective features given the objective ones from the isolated observations.
This step corresponds to automating the review process of coffee beans on those coffee beans for which only the objective features are available.
Since the predicted subjective features are in the $(0, 10]$ range interval, we experimented with three different models that are suitable for multi-target regression, i.e., RFs, SVR and a MLP. 
A thorough evaluation of these models to find the best one for our application will be presented in Section~\ref{sec:ob2sub}, where the average RMSE for each attribute was selected as the metric for evaluation.

Following the prediction step, both the predicted and the human-elaborated subjective features of the coffee beans in our dataset are combined together in the recommendation space.
Using a simple yet effective approach -- the k-nearest neighbour (kNN), our model recommends coffee beans by matching users' preferences in the recommendation space.

\subsection{Integration with the robot}
\label{ssec:introbot}

Our coffee bean recommender system would be deployed on a humanoid robot such as Pepper\footnote{https://www.softbankrobotics.com/emea/en/pepper}. Using a graphical user interface with visual sliders on Pepper's tablet, the user would choose how they would like their coffee, with varying degrees of sweetness and acidity, different aromas, flavours and aftertastes. After passing these through the coffee bean recommendation model discussed in the previous section, a list of coffee recommendations would be shown on Pepper's tablet. In reality, our system would also work on a tablet, and integration with the robot is not strictly necessary for the actual implementation. However, using a robot to complete such a task is important to enhance the user experience by introducing a service robot to help customers in a more natural way. Perspectives and analyses on implementing humanoid robot applications as interactive systems for customer engagement have been significantly tested in different scenarios such as the hospitality industry \cite{robot2hotel}. It has been repeatedly shown that human engagement tends to be more evident when an anthropomorphic robot interacts with the customer \cite{10.1145/3371382.3378343}. One such study has shown that people interacting with an emphatic robot showed a higher attention to the task amongst people who interacted with a static robot (or non-engaging devices) \cite{10.1145/3371382.3378343}. Here, the authors believe that an emphatic robot creates an affective bond between the robot and the user, which keeps the user engaged for a longer period of time. 
These results motivated our choice for using a robot instead of a tablet to deploy our coffee bean recommendation system.

\section{Experimental evaluation}
\label{sec:experimentalevaluation}
To validate our approach, we propose an experimental setup where we evaluated different stages of the overall system. 
We carried out quantitative experiments on the subjective features prediction model and the overall recommendation system.

\subsection{Subjective features prediction model}
\label{sec:ob2sub}
The first stage comprises the regression model for predicting the subjective features. 
As outlined in the previous section, RF, SVR and MLP were used as regression models for comparison. 
The results for each regressor are cross validated (with 10-folds) using the RMSE as performance metric (the lower the better).
For the RF, twenty decision trees were generated. 
For the SVR, a radial basis was used as kernel, with the penalty and gamma parameters having different values for each subjective feature. 
We used random search as the hyper-parameter optimisation strategy to find the optimal architecture and configuration of our MLP.
The resulting model consists of three fully connected hidden layers with 256 neurons each, and rectified linear units \cite{nair2010rectified} as activation function. 
We also used a dropout mask \cite{srivastava2014dropout} between successive layers to prevent overfitting and improve the generalisation capabilities of the model. 
The MLP was trained to minimise the RMSE of the model's predictions with the Adam optimiser \cite{kingma2014adam}.

As illustrated by Table~\ref{tab:ob2subresults}, the average RMSE for the RF across all attributes was 0.3562. 
For the best MLP configuration, the average RMSE achieved was 0.3580. 
With the SVR, an average RMSE across all attributes of 0.3750 was obtained. 
As the lowest RMSE attained was for the RF, this was chosen as the best model for our system.

\subsection{Recommender system}

To evaluate the recommender system as a whole, we measured the accuracy of the recommendations. 
In particular, we fitted a kernel density estimator on the subjective features of our dataset from which we sampled 100 new observations in order to simulate users' inputs from Pepper.
First, we run the kNN to identify the $k$ most similar coffee beans to the users' inputs. 
Since we expect our system to be deployed in a transductive setting, where the quality measures of some coffee beans are not available, we repeat the same procedure on a processed version of the full dataset. 
This is realised by considering a partition of it as unknown, so that we can use the regression model to obtain the subjective features needed by the kNN algorithm.
In this way, we can compare the coffee recommendations from the kNN by using the full dataset (i.e. the groundtruth) with those obtained by considering a partition from the predictions, to see how the latter ones diverge from the former. 
This allows to compute a measure of recommendation accuracy as the cardinality of the intersection between the recommendations from the full dataset and those from the dataset with predicted features. 
More formally, given a user's input $\mathbf{u} = [u_1, u_2, \dots, u_8]$ where $u_i$ is a subjective feature in Table \ref{tab:features}, we denote the groundtruth recommendations from the kNN on the full dataset $D$ as $rec_{ground}(\mathbf{u}, D^{full}) = [r_1, r_2, \dots, r_k]$, and the recommendations from the kNN on the dataset with $m \times |D|$ ($0 \leq m \leq 1)$ observations with unknown subjective features as $rec_{pred}(\mathbf{u}, D^{m}) = [p_1, p_2, \dots, p_k]$. 
The recommendation accuracy for the user input $\mathbf{u}$ is computed as:

\begin{equation}
    rec\_acc(\mathbf{u}) = \frac{|rec_{ground} \cap rec_{pred}|}{k}
\end{equation}

As reported in Table \ref{tab:recsys-eval}, we tested different values of $m$, and used cross-validation to obtain a more reliable estimate of the recommendation accuracy.
This ensures that our results are not biased by the dataset partitioning.

\begin{table}[!t]
\centering
\label{tab:ob2subresults}
\caption{The results tabulated illustrate the average RMSE for the seven predicted subjective features.}
\begin{tabular}{|c|c|}
\hline
\textbf{{ Model}} & \textbf{{Average RMSE}}\\\hline
\textit{\textbf{Random Forest (RF)}} & \textit{\textbf{0.3562}}\\\hline
Multilayer Perceptron (MLP) & 0.3580\\\hline
Support Vector Regression (SVR) & 0.3750\\\hline
\end{tabular} \\
\end{table}

\begin{table}[!t]
\centering
\label{tab:recsys-eval}
\caption{Recommendation accuracy with different sizes of the dataset partition for which subjective features have to be predicted.}
\begin{tabular}{|c|c|}
\hline

\textbf{{Prediction Size}} & \textbf{{Recommendation Accuracy}} \\ \hline
10\%            & 0.927000 \\ \hline
20\%            & 0.860600 \\ \hline
33\%            & 0.778667 \\ \hline
50\%            & 0.695500 \\ \hline
\end{tabular} \\
\end{table}

\section{Conclusion and future work}
\label{sec:conclusion}

In our work, we proposed a coffee bean recommendation system which to the best of our knowledge has not been addressed before. Whilst human coffee connoisseurs excel at such a task, modelling this recommender trait using artificial intelligence is non-trivial. We showed how supervised and unsupervised models can be applied for coffee bean recommendation. The aim of this work is to define regression models which can be used to predict subjective attributes, given objective ones. Using the random forests model, we obtained a RMSE of 0.3562. We also combined this with the k-nearest neighbours method to identify the most similar coffee beans to the users' inputs. In conjunction, we proposed an overview of how our model can be deployed on a service robot. 
As future work, we plan to deploy our model on a real robot and study how users would interact with the system. 

Specifically, an exciting research direction would be to study the role the robot would play to improve human-robot interaction and whether using a personalised coffee bean recommendation system would encourage return customers. Moreover, given the COVID-19 pandemic we have been experiencing, we would be interested to study whether using humanoid robots versus humans to recommend coffee would attract more customers as they would perceive the environment to be safer with less human interaction. We believe our presented work is a promising step in the direction of having service robots providing more personalised human-robot interaction.

\section*{Acknowledgment}
The authors would like to thank Prof. Gavin Brown for supporting this project through constructive feedback.



%
\bibliographystyle{IEEEtran}
\bibliography{cfpaperrefs}

\end{document}